\documentclass{jfm}
\usepackage{graphicx}
\usepackage{newtxtext}
\usepackage{newtxmath}
\usepackage{natbib}
\usepackage{hyperref}
\hypersetup{
    colorlinks = true,
    urlcolor   = blue,
    citecolor  = black,
}
\newcommand{\RomanNumeralCaps}[1]
\linenumbers

\title{Stable and unstable supersonic stagnation of an axisymmetric rotating magnetized plasma}
\author{Andrey Beresnyak\aff{1}
  \corresp{\email{andrey.beresnyak@nrl.navy.mil}},
  Alexander L. Velikovich\aff{1},
  John L. Giuliani\aff{12},
  \and Arati Dasgupta\aff{1}}

\affiliation{\aff{1}Naval Research Laboratory, Washington, DC
\aff{2}Emeritus}

\begin{document}
\maketitle

\begin{abstract}
The Naval Research Laboratory ``Mag Noh problem'', described in this paper, is a self-similar magnetized implosion flow, which contains a fast MHD outward propagating shock of constant velocity. We generalize the classic \citet{Noh1983} problem to include azimuthal and axial magnetic fields as well as rotation. Our family of ideal MHD solutions is five-parametric, each solution having its own self-similarity index, gas gamma, magnetization, the ratio of axial to the azimuthal field, and rotation. While the classic Noh problem must have a supersonic implosion velocity to create a shock, our solutions have an interesting three-parametric special case with zero initial velocity in which magnetic tension, instead of implosion flow, creates the shock at $t=0+$.
Our self-similar solutions are indeed realized when we solve the initial value MHD problem with finite volume MHD code Athena. We numerically investigated the stability of these solutions and found both stable and unstable regions in parameter space. Stable solutions can be used to test the accuracy of numerical codes. Unstable solutions have also been widely used to test how codes reproduce linear growth, transition to turbulence, and the practically important effects of mixing. Now we offer a family of unstable solutions featuring all three elements relevant to magnetically driven implosions: convergent flow, magnetic field, and a shock wave.
\end{abstract}

\begin{keywords}
MHD, Plasmas, Shock waves
\end{keywords}

\section{Introduction}
Fast magnetically-driven implosions of axisymmetric plasma columns and metal liners are essential for many applications of high-energy-density physics (HEDP) and inertial confinement fusion (ICF). Z-pinch implosions are used to generate high x-ray 
and neutron yields; see \citet{Pereira1988,Giuliani2015,Jones2006,Coverdale2007,Ampleford2014,Sinars2020}, and references therein. 
\begin{figure}
\begin{center}
\includegraphics[width=0.4\columnwidth]{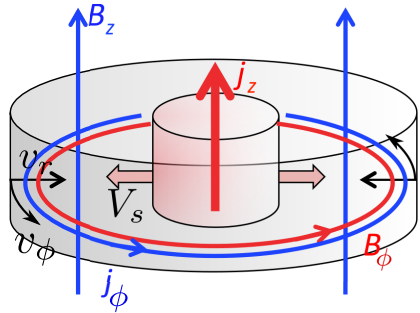}
\end{center}
\caption{A drawing of a supersonic, shocked, axisymmetric, magnetized, rotating implosion. Here $v_r$ is an implosion velocity, $v_\phi$ -- rotation velocity, $V_s$ -- expanding shock velocity, $B_\phi$ and $B_z$ are the azimuthal and axial magnetic field produced by, correspondingly, $j_z$ and $j_\phi$ current densities.}
\label{impl_drawing}
\end{figure}
The implosion of a thin metal liner compressing a laser-preheated plasma is the critical component of the Magnetized Liner Inertial Confinement Fusion (MagLIF) approach to the ICF; see \citet{Slutz2010,Slutz2012,Cuneo2012,Gomez2020}. The plasma and metal loads are imploded by the azimuthal magnetic field of the axial current passed through the load. In many cases, an external axial magnetic field is applied to stabilize the implosion or provide the ``magneto-thermo-insulation'' \citep{Lindemuth1983} of the stagnated plasma, which is the critical issue for all approaches to magnetized target fusion and magneto-inertial fusion \citep{Turchi2008,Garanin2013,Wurden2016,Lindemuth2017}, including the MagLIF. The axial magnetic field, in turn, can affect the implosion dynamics in various ways, not all of which are fully understood; see \citet{Felber1988,Shishlov2006,Rousskikh2017,Mikitchuk2019,Conti2020,Seyler2020a}, and references therein. Recent experiments with gas-puff Z-pinch implosions in an external magnetic field \citep{Cvejic2021} demonstrated a self-generated rotation of the imploding axisymmetric plasma.
The LINUS project pursued at NRL since the early 1970s, the earliest approach to magnetized target fusion (see \citet{Robson1973,Turchi1980} and references therein) involved both magnetic flux compression of the fusion plasma and rotation of the liner compressing it. Here, the rotation stabilized the liner implosion against the Rayleigh-Taylor instability by reversing the effective gravity at its inner surface. The papers \citet{Book1974,Barcilon1974,Book1979} predicted the stabilization theoretically, and
\citet{Turchi1976,Turchi1980} demonstrated it experimentally. A modified version of this approach is presently pursued by General Fusion, see \citet{Huneault2019} and references therein. The studies of the 1970s revealed that compressibility of the rotating liquid liner and formation of shock waves in it are not negligible, see the discussion by \citet{Book1979}. They are more relevant and can become dominant if the compressed rotating magnetized fluid is a low-density plasma, as in Z-pinch implosions.
A prominent feature of fast Z-pinches is the thermalization of the imploding plasma's kinetic energy at stagnation near the axis via an expanding accretion shock wave. The kinetic-to-thermal energy conversion in an expanding cylindrical shock wave plays a significant, maybe dominant role in the keV X-ray radiation production in both gas-puff and wire-array Z-pinches \citep{Basko2012,Maron2013}. The radial flow of the low-density, cold plasma ablated from the wires in cylindrical wire-array Z-pinch implosions stagnates via an expanding shock wave into a precursor plasma column. This process strongly affects the dynamics and stability of wire-array implosions; see \citet{Lebedev2001,Bott2009}, and references therein.

A detailed description of the supersonic stagnation process, even for the one-dimensional (1D) axisymmetric case, requires MHD numerical simulation. It would be helpful to gain insight into the dynamics and stability of the implosion and stagnation and to enable verification of the codes used for this purpose by developing exact analytical or semi-analytical solutions. To do this, we can use the Lie symmetries of the governing non-stationary ideal-MHD equations \citep{Ovsiannikov1982,Galas1993}, reducing them to ordinary differential equations and constructing their exact self-similar solutions. The presence of the magnetic field significantly constrains the MHD families of self-similar solutions compared to their counterparts in compressible gas dynamics. Well-known examples of exact gas-dynamic solutions describing supersonic stagnation are the solution of collapsing-shock or collapsing-cavity problems \citep{Guderley1942,Hunter1960} extended to the post-collapse reflected-shock stage \citep{Stanyukovich1960,Zeldovich2002,Lazarus1981}. Axisymmetric MHD solutions of this kind with an azimuthal magnetic field are harder to construct. They typically involve singular features, such as a time-dependent line current on-axis \citep{Liberman1986,Venkatesan1992,Lock2008}. Finding semi-analytic solutions suitable for comparison to direct numerical simulations, often requires going beyond self-similarity, using, for example, the extension of the geometric shock dynamics approximation \citep{Whitham2011} as done by \citet{Pullin2014} and \citet{Mostert2016}. For the situations when exact self-similar solutions can be constructed, they do not appear suitable for code verification: ``Numerical calculations stubbornly refuse to exhibit attraction to a self-similar asymptotic state,'' as noted by \citet{Lock2008}.
Nevertheless, it is possible to describe a supersonic axisymmetric stagnation of magnetically driven imploding mass by exact self-similar solutions of ideal MHD equations. Such solutions represent the flow stage after the diverging reflected MHD shock starts propagating into the converging magnetized plasma, leaving behind the stagnated plasma. Formally, they are MHD generalizations of the well-known Noh solution for a constant velocity gas that stagnates against an axis of symmetry, producing an outward moving shock. In gas dynamics, solutions of this class were discovered by Sedov and published in 1954 in the third Russian edition of his book, see \citet{Sedov1959}, Chapter 4, Section 7 ``The problem of an implosion and explosion at a point.'' They did not attract attention until being rediscovered by \citet{Noh1983,Noh1987}. The simple, explicit Noh solution obtained for an infinitely strong shock in an ideal gas became the workhorse of compressible hydrocode verification for over three decades. It can be generalized for an arbitrary shock strength \citep{Velikovich2018} and equation of state \citep{Velikovich2018a}. An MHD generalization for the Z-pinch geometry, i. e., with the azimuthal magnetic field, is also available \citep{Velikovich2012}. This 1D self-similar solution has been successfully used to verify several hydrocodes, including Mach2, Athena++, FLASH \citep{Tzeferacos2012}. The consistent convergence of two-dimensional (2D) numerical solutions obtained with MHD hydrocodes to the 1D self-similar solution illustrated the stability of the latter.

In what follows, we present an extended family of self-similar solutions of cylindrical, magnetized Noh problems, the magnetic fields of which are both azimuthal and axial. It also includes sheared rotation of the stagnating plasma. There is a clear analogy between the dynamic effects of the axial magnetic flux compression and the rotation of an imploded plasma column. The force impeding implosion is inversely proportional to the cubed radius of the column, as explained by \citet{Velikovich1995} for the case of solid rotation and homogeneous-deformation self-similar solutions. Of course, the analogy remains valid for a broader class of solutions described here. The actual production of sheared rotation in Z-pinch implosions in an axial magnetic field is a very recent experimental finding \citep{Cvejic2021}.  The new solutions will be helpful both to analyze the experimental results on the self-generated rotation of magnetized Z-pinch plasmas and to verify MHD codes aiming to simulate this effect. We also imagine it could be useful in the astrophysical community dealing with shocked accretion from magnetized and rotating accretion disks. 

Instead of an elaborate semi-analytic procedure used by \citet{Velikovich2012} to construct a solution to be tabulated, we present its implementation as a Python code that the readers can use to generate any solution of this family. This option makes it more accessible for analysis and code verification than the solutions that describe axisymmetric MHD shock propagation through a rotating medium in astrophysical literature; see \citet{Vishwakarma2007,Nath2020}, and references therein. 
Our new analysis reveals a previously unknown property of magnetized Noh solutions, in contrast with the classical solution \citep{Noh1983,Noh1987} and its gas-dynamic generalizations. Specifically, a stagnating flow with a radially expanding accretion shock emerges at $t=0+$ even if the plasma is initially at rest everywhere. The stagnation is caused by the imbalance of radial forces acting upon the plasma, and it takes place with or without the axial magnetic field and/or rotation. We have demonstrated a consistent convergence of Athena++ 1D simulation results to our self-similar solutions for all cases: with or without the axial magnetic field, initial rotation, and initial radial motion.
Another new feature discovered in our 2D Athena++ numerical simulations of supersonic stagnation in an axial magnetic field is the instability of this flow obtained for some perturbation modes with specific base-flow parameters. This behavior was unexpected because gas-dynamic Noh-type accretion-shock flows are typically stable, cf. \citet{Velikovich2016,Huete2021}. 2D simulations of stagnation described by the magnetized Noh solution with an azimuthal magnetic field detected no instability. The instability emerges only when the axial magnetic field is added. It is not of magneto-rotational origin because it manifests itself even without rotation. While the physical mechanism driving this new instability deserves further study, we found parameter ranges in which our new solutions are stable and can be confidently used for code verification in 2D.

The paper is structured as follows: Figure \ref{impl_drawing} shows our problem setup. In Section 2 we present self-contained derivation of the main dynamical equations. In Section 3 we describe shock jump conditions as well as a general procedure of obtaining the self-similar solution and its verification with MHD code Athena. In Section 4 we consider an interesting special case with zero initial velocity. In Section 5 we describe rotating solutions and their verification. In Section 6 we describe our numerical findings of stability of our solutions to azimuthal perturbations. In Section 7 we provide some examples of code verification on a 2D Cartesian grid. In Section 8 we summarize and discuss our findings. In Section 9 we refer the reader to our publicly released code. Appendix A and B are dedicated to deriving asymptotic behavior of our solutions close to the origin and at very large distances from the origin.

\section{Analytic derivation}
\label{analytic}
We start with the equations for ideal, isentropic, cylindrically symmetric MHD flow. For the density $\rho$, radial velocity $v_r$, azimuthal velocity $v_\phi$, pressure $p$, azimuthal magnetic field $B_\phi$, axial magnetic field $B_z$ we write the continuity, radial momentum, angular momentum, specific entropy and Faraday's law as
\begin{equation}
\frac{\partial \rho}{\partial t}+\frac{1}{r} \frac{\partial}{\partial r}(r \rho v_r)=0,
\end{equation}
\begin{equation}
\frac{\partial(\rho v_r)}{\partial t}+\frac{1}{r} \frac{\partial}{\partial r}\left(r \rho v_r^{2}\right)+\frac{\partial p}{\partial r}+\frac{\partial}{\partial r}\left(\frac{B^{2}_\phi+B^{2}_z}{8 \pi}\right)+\frac{B^{2}_\phi}{4 \pi r}-\frac{\rho v^2_\phi}{r}=0, 
\end{equation}
\begin{equation}
\frac{\partial(\rho v_\phi)}{\partial t}+\frac{\partial}{\partial r}\left(\rho v_r v_\phi\right)+
2\frac{\rho v_r v_\phi}{r}=0, 
\end{equation}
\begin{equation}
\frac{\partial p}{\partial t}+\frac{\gamma p}{r} \frac{\partial}{\partial r}(rv_r)+v_r \frac{\partial p}{\partial r}=0.
\end{equation}
\begin{equation}
\frac{\partial B_\phi}{\partial t}+\frac{\partial}{\partial r}(v_r B_\phi)=0
\end{equation}
\begin{equation}
\frac{\partial B_z}{\partial t}+\frac1r\frac{\partial}{\partial r}(r v_r B_z)=0
\end{equation}
In the case of cylindrical symmetry, the radial component of the magnetic field, $B_r$ is zero due
to divergence-free condition for the magnetic field. We will assume that the z-velocity component is zero
because, due to Galilean invariance it does not participate in the dynamics, but is rather passively advected by the flow. Later in the paper, we will study the stability of this problem with respect to the
azimuthal perturbations, in which case most components will be non-zero. However, while doing this, we will keep the origin at rest, which is implied by our cylindrical coordinates.

We use the Eqs. above to describe a self-similar imploding
cylindrical plasma, which stagnates in a shock front expanding at a
constant velocity. We assume that at the shock the entropy conservation should be replaced with
a shock jump condition, as described below. We define the self-similar variable as
\begin{equation}
\eta \equiv \frac{r}{t} \Rightarrow r=\eta t,
\end{equation}
we also assume without loss of generality that the outward propagating shock velocity is unity, i.e., the shock front is always located at $\eta=1$. We seek the self-similar MHD solution in the form
\begin{eqnarray}
v_r(r,t)&=&\eta U(\eta) \\
v_\phi(r,t)&=&\eta W(\eta)  \\
\rho(r, t)&=&t^{2 \chi} N(\eta) \\
p(r,t)&=&t^{2 \chi} P(\eta)\\
B_\phi(r,t)&=&t^{\chi} \sqrt{4\pi} H_\phi(\eta)\\
B_z(r,t)&=&t^{\chi} \sqrt{4\pi} H_z(\eta)
\end{eqnarray}
Here $\chi>0$ is a dimensionless parameter characterizing the self-similar scaling
of physical quantities. The initial conditions for our problem will be
\begin{eqnarray}
v_r(r, t=0)&=&v_{r0} ; \quad v_\phi(r, t=0)=v_{\phi 0} ; \quad \rho(r, t=0)=\rho_{0} r^{2 \chi} \nonumber \\
B_{\phi,z}(r, t=0)&=&B_{\{\phi,z\} 0} r^{\chi} ; \quad p(r,t=0)=0,\label{init}
\end{eqnarray}
we also restrict ourselves to zero initial pressure, as we seek to generalize the Noh problem. With finite
density this corresponds to zero temperature of inflowing gas.
These initial values are related to the limit $\eta\to\infty$ as
\begin{eqnarray}
v_{r0}&=&\lim_{\eta\to\infty} U(\eta) \eta,\quad 
v_{\phi 0}=\lim_{\eta\to\infty} W(\eta) \eta,\nonumber \\
\rho_0&=&\lim_{\eta\to\infty} N(\eta) \eta^{-2\chi},\quad 
B_{\{\phi,z\} 0}=\sqrt{4\pi} \lim_{\eta\to\infty} H_{\phi,z}(\eta) \eta^{-\chi}, \label{NBH_init}
\end{eqnarray}
To transform variables $(r,t) \to (\eta, t)$, we use
\begin{equation}
\frac{\partial}{\partial r} \rightarrow \frac{1}{t} \frac{\partial}{\partial \eta}, \quad \frac{\partial}{\partial t} \rightarrow \frac{\partial}{\partial t}-\frac{\eta}{t} \frac{\partial}{\partial \eta};
\end{equation}
Then, in the MHD equations the time derivative drops out and we arrive at
\begin{eqnarray}
(U-1) \frac{d \ln N}{d \ln \eta}+\frac{d U}{d \ln \eta}+2 U+2 \chi=0 \label{ssim_rho}\\
(U-1)\left[\frac{d U}{d \ln \eta}+U\right]-W^2+\frac{1}{\eta^{2} N}\left[\frac{d P}{d \ln \eta}+\frac{H_\phi}{\eta} \frac{d(\eta H_\phi)}{d \ln \eta}+\frac12\frac{dH_z^2}{d \ln \eta}\right]=0 \\
(U-1) \frac{d \ln W}{d \ln \eta}+2U-1=0\\
(U-1) \frac{d \ln P}{d \ln \eta}+\gamma \frac{d U}{d \ln \eta}+2 \gamma U+2 \chi=0 \label{ssim_p}\\
(U-1) \frac{d \ln H_\phi}{d \ln \eta}+\frac{d U}{d \ln \eta}+U+\chi=0 \label{ssim_hphi}\\
(U-1) \frac{d \ln H_z}{d \ln \eta}+\frac{d U}{d \ln \eta}+2U+\chi=0 \label{ssim_hz}
\end{eqnarray}
Note the following symmetry of Eqs.~(\ref{ssim_rho})-(\ref{ssim_hz}): they are invariant under
arbitrary scaling transforms of $P$, $N$ and $H$, provided that $P/\eta^2N$, $H_\phi^2/\eta^2N$, $H_z^2/\eta^2N$
stay constant. Physically this means rescaling units for pressure, density, and magnetic field while keeping sonic and Alfv\'en speed the same. This symmetry can be utilized to simplify the system (\ref{ssim_rho})-(\ref{ssim_hz}) further
by introducing new variables
\begin{equation}
S=\frac{\gamma P}{\eta^{2} N(1-U)} \quad {\rm and} \quad A_{\{\phi,z\}}=\frac{H^{2}_{\{\phi,z\}}}{\eta^{2} N(1-U)},
\end{equation}
as done in \citet{Liberman1986,Felber1988,Venkatesan1992}. We define Mach numbers as
\begin{equation}
M_{s}^{2}=\frac{(v_r-\eta)^{2}}{c_{s}^{2}}=\frac{1-U}{S}, \quad M_{A\{\phi,z\}}^{2}=\frac{(v_r-\eta)^{2}}{v_{A\{\phi,z\}}^{2}}=\frac{1-U}{A_{\{\phi,z\}}},
\label{M_def}
\end{equation}
where $c_{s}=\sqrt{\gamma p/ \rho}$ and $v_{A_{\{\phi,z\}}}=B_{\{\phi,z\}} / \sqrt{4 \pi \rho}$ are the local values of the speed of sound and Alfv\'en speed, $\eta$ is the ``phase'' velocity of the self-similar profile. These are the local Mach numbers of the radial flow velocity relative to the surfaces of constant $\eta$, i.e., $v_r-\eta=\eta(U(\eta)-1)$. At $\eta=1$, the Mach numbers (\ref{M_def}) characterize the flow velocity with respect to the shock front. It is also useful to introduce the fast magnetosonic Mach number as
\begin{equation}
M_F^{2}=\frac{1-U}{S+A_\phi+A_z}.
\end{equation}
Introducing new variables, we are left with five equations:
\begin{eqnarray}
(U+S+A_\phi+A_z-1) \frac{d U}{d \ln \eta}+U(U-1)-W^2 & &\nonumber \\
+2 S\left(U+\frac{\chi}{\gamma}\right)+(\chi+1)A_\phi+(\chi+2U)A_z&=&0 \label{U_main}\\
(U-1) \frac{d \ln W}{d \ln \eta}+2U-1&=&0\label{W_main}\\
(U-1) \frac{d \ln S}{d \ln \eta}+\gamma \frac{d U}{d \ln \eta}+2(\gamma U-1)&=&0\label{S_main} \\
(U-1) \frac{d \ln A_\phi}{d \ln \eta}+2 \frac{d U}{d \ln \eta}+2U-2&=&0\label{Aphi_main}\\
(U-1) \frac{d \ln A_z}{d \ln \eta}+2 \frac{d U}{d \ln \eta}+4U-2&=&0 \label{Az_main}
\end{eqnarray}
By combining equations (\ref{W_main})-(\ref{Az_main}) and (\ref{ssim_rho}) it is possible to write several explicit integrals of our system, namely
\begin{eqnarray}
S^{\chi+1}N^{1-\gamma}\left[\eta^2(1-U)\right]^{\gamma\chi+1}={\rm const},\label{S_int}\\	
A_\phi(1-U)^2\eta^2=v^2_{A\phi 0},\label{Aphi_int}\\	
A_z^{\chi+1}N^{-1}\left[\eta^2(1-U)\right]^{2\chi+1}=v^{2(\chi+1)}_{Az0}\rho_0^{-1},\label{Az_int}\\
W^2 A_z^{-1}(1-U)^{-2}=(v_{\phi 0}/v_{Az0})^2\label{W_int}.
\end{eqnarray}
Where $v_{A\{\phi,z\}0}=v_{A\{\phi,z\}}(t=0)$.	
These correspond to the frozen-in condition for specific entropy, $\phi$ and $z$ components of the magnetic field and conservation of angular momentum. Note that the first integral is broken at the shock, being
zero for $\eta>1$ and non-zero for $\eta<1$ because the specific entropy does not conserve in shocks. Its value in the post-shock fluid can be obtained from the shock jump condition that we describe in the next section. Numerically, it is often easier to solve the system (\ref{ssim_rho}) and (\ref{U_main})-(\ref{Az_main}) directly, without involving integrals.

We look for regular solutions which have $U\leq 0$, so $(U-1)$ is always non-zero.
The prefix $U+S+A_\phi+A_z-1$ in the Eq.~(\ref{U_main}) can be represented as $(U-1)(1-1/M_F^2)$.
The solution becomes singular at the fast magnetosonic point where the local velocity with respect
to the shock is equal to the group velocity of the fast magnetosonic mode. 
At large $\eta$ the fast magnetosonic Mach number $M_F$ is larger than unity, while
after crossing the fast shock it becomes smaller than unity. A regular solution is obtained when
the prefix in the Eq.~(\ref{U_main}) changes its sign only at the shock\footnote{The solution can,
in principle, pass fast magnetosonic point regularly, if we require that the last four terms of the Eq.~(\ref{U_main}) sum to zero at the fast magnetosonic point, but this will require one extra constraint imposed on the initial conditions. We will not seek these special
solutions and they will be considered elsewhere.}.

\section{Evaluating self-similar non-rotating solutions}
\begin{figure}
\includegraphics[width=1.0\columnwidth]{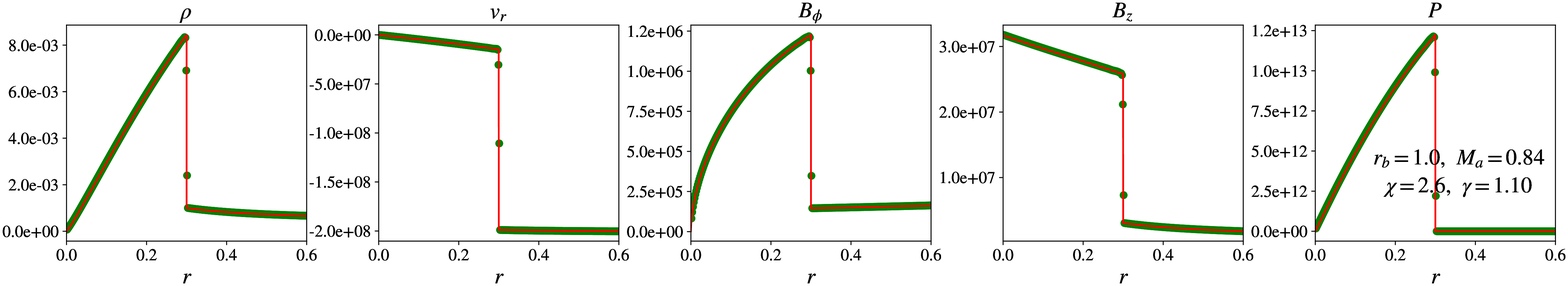}
%
\includegraphics[width=1.0\columnwidth]{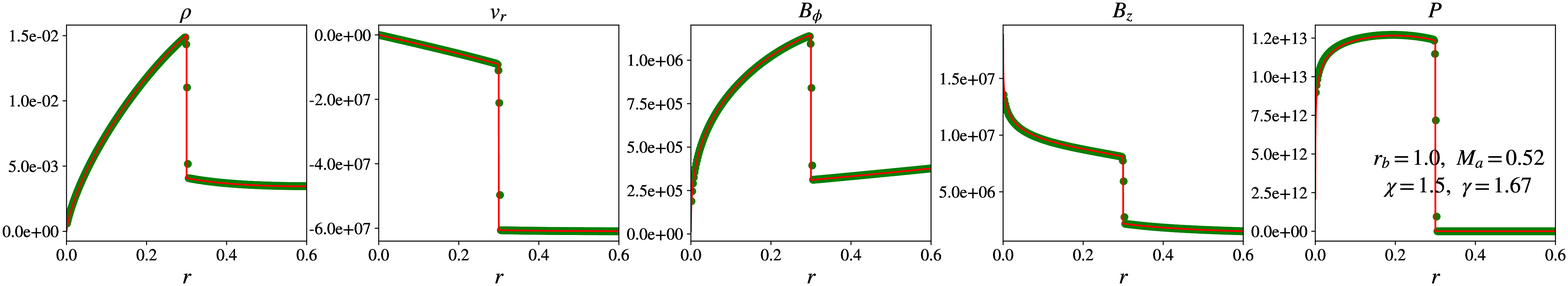}
\caption{Two example solutions of magnetized Noh (red) verified with Athena++ in one-dimensional cylindrical simulation (green circles). Both solutions initially have $B_z/B_\phi=1$. Top solution with $\chi=2.6, \gamma=1.1, v_A/V_s=0.84$ and $v_0/V_s=20$ have regular $j_\phi$ on axis, while bottom with $\chi=1.5, \gamma=5/3, v_A/V_s=0.52$ and $v_0/V_s=6.1$ have singular $j_\phi$ on axis. Note that Athena++ is an MHD code that solves initial value problem and is not in any way aware of either the solution self-similarity or the expected shock speed.}
\label{two_solutions}
\end{figure}

We solve the main system (\ref{U_main})-(\ref{Az_main}) in a natural fashion, starting with initial conditions
(\ref{init}) that impose boundary conditions for our dynamic variables at infinity (\ref{NBH_init}).
Numerically, we start with very large value of $\eta$ and integrate the system (\ref{U_main})-(\ref{Az_main})
towards $\eta=1$ where we assume the shock is located. We do so without a lack of generality, because
if the shock is located at a different $\eta$, i.e. the shock has a different velocity, we can rescale all quantities to renormalize back to the shock speed of unity. For example, we can keep $N$ constant and rescale initial $A_{\phi,z}$ by a shock speed squared. At large values of $\eta$ our numerical solution is consistent with the analytical asymptotic formulas presented in Appendix B.

At the shock we apply the standard MHD shock jump conditions \citep{ll08,Liberman1986}:
\begin{eqnarray}
U &\to& 1-\mu(1-U),\\
A_{\{\phi,z\}} &\to& A_{\{\phi,z\}}/\mu^2, \\
S &\to& (1-U)/M_S^2, 	
\end{eqnarray}
where the density compression parameter $\mu=\rho_1/\rho_2$ and $M_S$ are determined by the solution of a quadratic equation for $\mu$:
\begin{eqnarray}
\mu&=&\frac{\gamma_2+1/M_{A}^2+\sqrt{(\gamma_2+1/M_{A}^2)^2+4(1-2\gamma_2)(2-\gamma_2)/M_{A}^2}}{4-2\gamma_2},\\
\frac1{M_S^2}&=&\gamma\left(1-\mu-\frac{1/\mu^2-1}{2M_{A}^2}\right),
\end{eqnarray}

where $\gamma_2=1-1/\gamma$ and $M_A^{-2}=M_{A\phi}^{-2}+M_{Az}^{-2}$. Keep in mind that $M_S^2$ is different from $M_s^2$ in Eq.~\ref{M_def} in that the former is the ratio of pre-shocked $1-U$ to post-shock $S$.
After the shock, we continue to integrate the system (\ref{U_main})-(\ref{Az_main}) keeping eye on the sign of $U$
and the sign of $1-M_F$.  We want to obtain the regular solution that stagnates on-axis, i.e., have $\eta U \to 0$ as $\eta \to 0$. In the Appendix A, we analytically derive boundary conditions for $U$ at the origin, $U_0$. For $\gamma>2$, $U_0=-\chi/\gamma$, for $\gamma<2$ and non-rotating case, $U_0=-\chi/2$. For the rotating case the expression is more complicated and provided in the Appendix A.

We can always obtain the solution that stagnates at larger $\eta$ by choosing sufficiently small $-v_0$.
At the same time, if $-v_0$ is too large, we have the solution that either crosses the fast magnetosonic point or has $U-U_0<0$ at $\eta=0$. Between these two extreme cases is our desired solution, which has $U=U_0$ at the origin.

Numerically, we can not integrate down to $\eta=0$, so we choose some small but finite $\eta_{\rm min}$ to stop integration.
We seek $v_0$ that minimizes $U-U_0$ at $\eta_{\rm min}$ by the bisection method, assuming that crossing the fast magnetosonic point means that $-v_0$ is too large while stagnating before $\eta_{\rm min}$ means that $-v_0$ is too small. Once we obtained the converged solution, at small values of $\eta$, our numerical solution is consistent with the analytical asymptotic formulas presented in Appendix A.

Thus we obtained a self-similar solution that has two parameters, $\chi$, and $\gamma$, and two initial conditions for $A_{\{\phi,z\}}$ determined from (\ref{NBH_init}). The solution for velocity and magnetic field expressed in physical units,
have two more degrees of freedom --- we can rescale the flow and the shock velocity to an arbitrary value, which will rescale pressure by the same factor squared and we can rescale density and magnetic field while keeping $B^2/\rho$ constant.
The physical solution of magnetized Noh, therefore, may be classified by six parameters: $\chi$, $\gamma$, shock velocity, initial density, and the two components of the initial magnetic field. We can also obtain solutions with finite initial pressure, but will not analyze them here. In Appendix A we derive
asymptotic behavior of density, pressure, $B_\phi$, and $B_z$ on-axis. For the most physically relevant
case of $\gamma<2$, we find that density, pressure, and $B_\phi$ go to zero, while $B_z$ goes to a constant.
The azimuthal current density $j_\phi$ that creates $B_z$ is regular if $\gamma<3/2$ and $\chi>2/(3-2\gamma)$. Otherwise, $j_\phi$ is singular, and since $B_z$ is energetically a dominant
term on-axis this will create sizable dissipation in non-ideal plasmas.

We found it is convenient to classify Mag Noh problems without rotation by four dimensionless numbers: the ratio of the initial axial to the initial azimuthal field $r_b$; initial Alfv\'en speed in units of the shock speed $M_a=v_A/V_s$ and $\chi$ and $\gamma$. The rotation adds another dimensionless parameter, which is a $v_{\phi 0}/v_{r0}$ ratio. 
\begin{figure}
\begin{center}
\includegraphics[width=0.85\columnwidth]{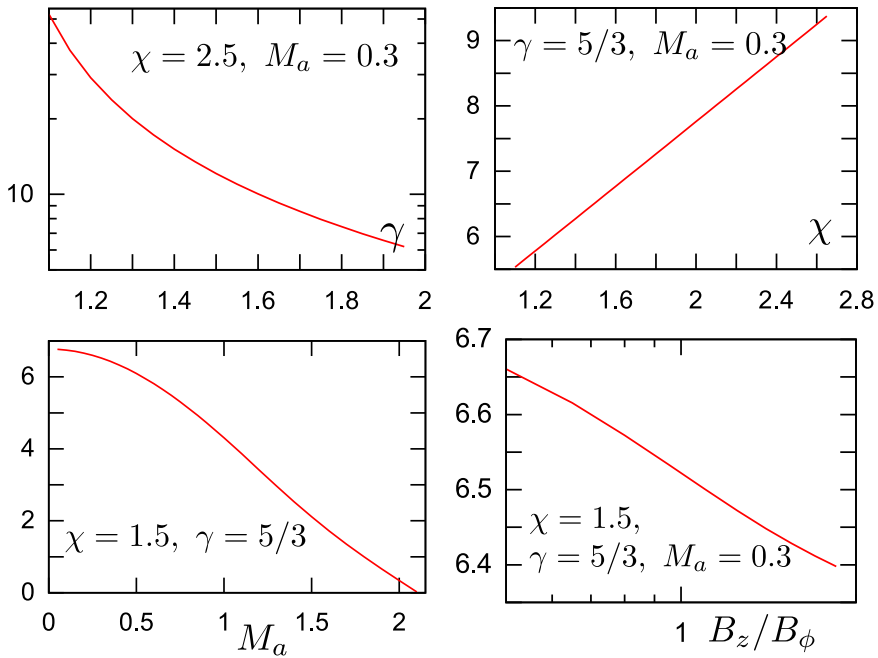}
\end{center}
\caption{Dependence of $-v_0/V_s$ vs. various parameters of the Mag Noh problem, in all but the last plot $B_z/B_\phi=1$. Also note that in the upper left the $v_0$ axis is logarithmic.}
\label{dependence}
\end{figure}

For a given set of parameters, we calculated the self-similar solution in a manner described above with a simple Euler ODE solver, starting at $\eta_{\rm max}=10^5$ and ending at $\eta_{\rm min}=10^{-5}$. Typically,
$2\times 10^5$ steps was enough to produce an accurate solution. We used the Athena code \citep{Stone2020,athena} in cylindrical geometry to verify that our solutions are realized in numerical calculation. Athena is a finite volume Godunov code and we used its one-dimensional solver with first-order spatial reconstruction\footnote{Higher order reconstruction schemes often produce oscillations next to the shock. In this work we did not focus on the fast convergence of the numerical solution and for the sake of simplicity we used first order. Our publicly released code can be trivially modified to test
higher order convergence.}. In Athena simulation we typically assume the shock speed of $10^7$cm/s, domain size of 8 cm with 4000 grid points and we evolve the solution to 30 ns, so that the shock is located at 0.3 cm. We use initial conditions (\ref{init}), reflecting boundary at the origin and inflow boundary at outer radius. The inflow boundary is not the correct boundary for our problem, but we assume that the domain size of 8 cm is large enough so that the perturbation set by the boundary do not propagate into the useful solution range 0-0.6 cm in 30 ns. Larger
domain sizes may be needed for a large inflow velocity. Simulation parameters are tunable and we release
our code to the public for an easy adoption by the community, see Section~\ref{release}.
Fig.~\ref{two_solutions} presents two example solutions, one of them having a regular and the other singular $B_z$ on-axis. We also release this verification code to the public (see Data Release).
We scanned our four-dimensional parameter space to determine the initial inflow speed normalized by the shock speed $v_0/V_s$ as a function of the four main parameters: $\chi, \gamma, M_a$, and $B_z/B_\phi$. The results are present in Fig.~\ref{dependence}. The strongest dependence is on $\gamma$, with $\gamma$
approaching unity the compressibility increases and the shock Mach number can be very high. The dependence on the ratio of the magnetic components is rather weak, similarly, the index of similarity
only moderately affects the strength of the shock. The dependence on the Alfv\'en Mach number upstream
is interesting in that it reveals a solution that has $v_0=0$. This is a rather interesting case of magnetic Noh because unlike classic Noh it does not have a flow hitting the wall, the initial condition is stationary.

\section{Stationary initial conditions}
\begin{figure}
\begin{center}
\includegraphics[width=1.0\columnwidth]{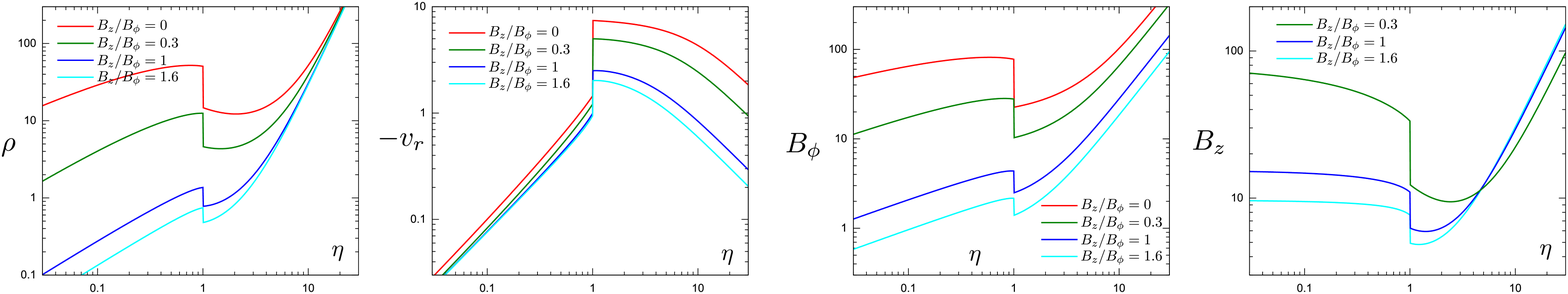}
\includegraphics[width=1.0\columnwidth]{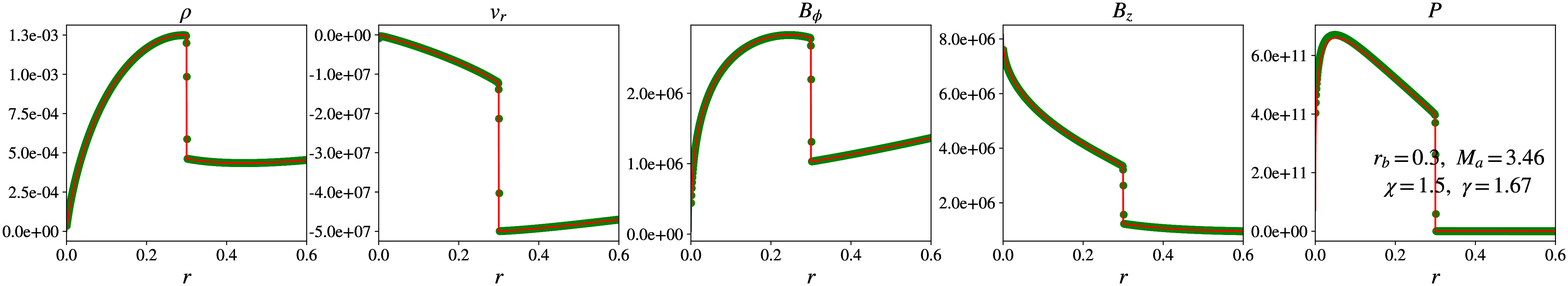}
\end{center}
\caption{Solutions with zero initial velocity with $\chi=1.5$ and $\gamma=5/3$. Top: several solutions with different ratio $B_z/B_\phi$. Bottom: Verification with Athena++ of the above $B_z/B_\phi=0.3$ solution with  $v_A/V_s=3.45$, singular $j_\phi$ on axis, self-similar (red) overplotted on Athena++ (green circles). Note that top plot is log-scale, while bottom is linear scale.}
\label{v0_sol}
\end{figure}
Let us look at this interesting special case with zero initial velocity in more detail. This case deserves closer scrutiny because it is closer to practical applications such as Z-pinches, where the gas, initially at rest, is driven towards the axis and is shocked and heated. As far as code verifications are concerned, stationary initial conditions are easier to set up in some codes, e.g., the moving mesh codes.

In the case of fluid at rest initially, the flow develops due to the magnetic tension. One may expect that it would take some time to accelerate the flow before it develops a shock, this is not the case with our self-similar solution, however.
In our solution the shock is present for all times $t>0$, this is perhaps because density is very small near the origin and it takes very small tension to create a shock. In order to obtain these solutions, we modified the solver described above. Now we do not change the initial velocity to find the stagnating solution, but rather we change $B_\phi$. For sufficiently small $B_\phi$ we expect the solution to stagnate before origin, while at sufficiently large $B_\phi$ the solution either hits the origin with non-zero velocity or passes the fast magnetosonic point. Again, we seek $B_\phi$ that minimizes $|U-U_0|$ at $\eta_{\rm min}$ by the bisection method. All solutions we found have a sizable $v_A/V_s$, between 2.0 and 4.8, which is not surprising since some minimum magnetic tension is needed to create a shock. 

While studying the parameter space we found that solutions with different $\chi$ and $\gamma$ look very similar. Changing the $B_z/B_\phi$ ratio, however, makes quite a bit of difference. In Figure~\ref{v0_sol} we show several solutions for $\chi=1.5$ and $\gamma=5/3$ and different field ratios. Generally, the lower $B_z$ results in a stronger shock with larger compression of the gas and both components of the field. The smaller the $B_z/B_\phi$ ratio is, the more singular $j_\phi$ becomes at the origin.
In Figure~\ref{v0_sol}, bottom, we show the verification of the $B_z/B_\phi=0.3$ solution with Athena++. Note that close to the origin the gas pressure goes to zero and is replaced by the magnetic pressure of $B_z$. The field ratio $B_z/B_\phi$ is usually small in realistic Z-pinches and such solutions are potentially of practical interest since singular $j_\phi$ can provide extra heating near the low-density origin and create a hot spot. Such heating is, of course, not present in our solutions of ideal MHD.

\section{Rotating solutions}
\begin{figure}
\begin{center}
\includegraphics[width=0.95\columnwidth]{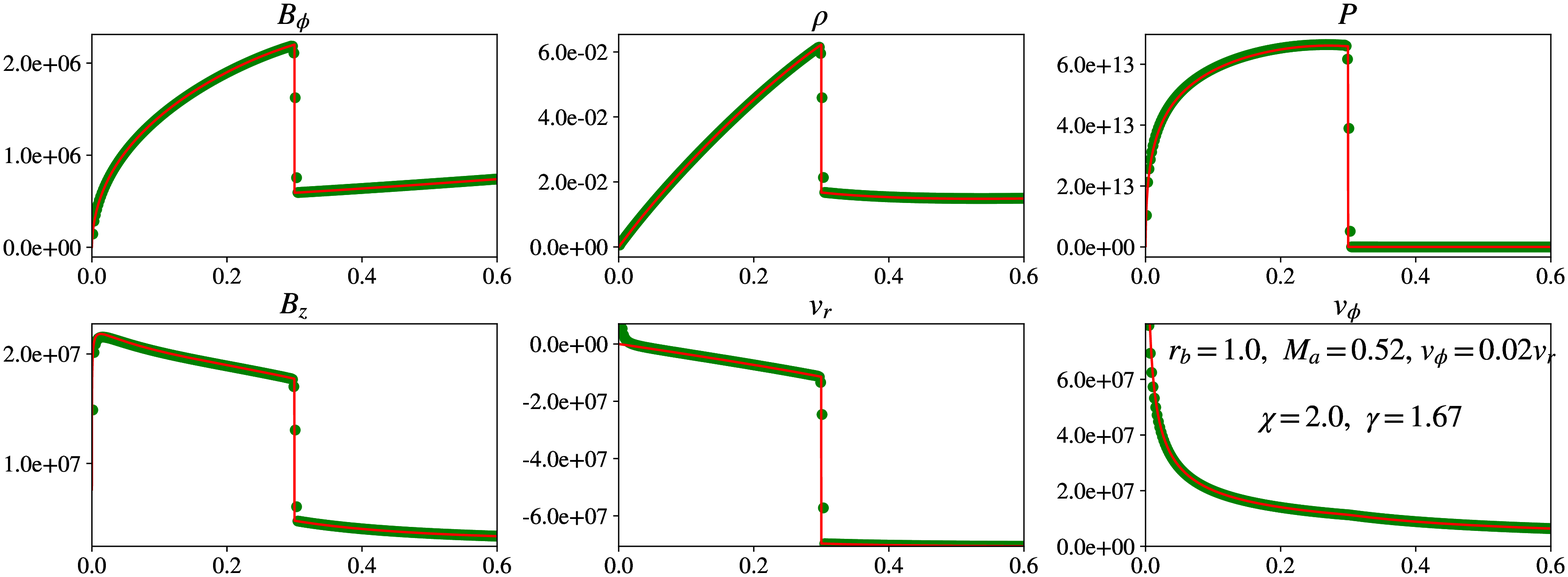}
\end{center}
\caption{Rotating solution produced by self-similar solver (red) overplotted on the solution produced by Athena++(green dots).}
\label{rotating}
\end{figure}
We can obtain rotating solutions as well, in a manner similar to described above. For convenience, we introduce the rotation parameter, which is the ratio of initial rotation velocity to radial velocity. Rotating initial conditions are unusual in that we have to set rotating velocity to constant, even at the origin, which produces infinite vorticity. Our self-similar solver, however, does not have an issue with that, because initial conditions are set at infinity. It is, however, very interesting how numerical codes will handle such a situation.
Fig.~\ref{rotating} shows verification of our self-similar solution with the solution produced by Athena++ and the agreement is pretty good. One feature of the rotating solution is a singularity of rotation velocity at the origin, which is a consequence of the conservation of angular momentum. The other interesting feature is that adding rotation makes $B_z$ go to zero on-axis. This is easy to understand as the radial electric field $E_r$, which is absent in non-rotating solutions is now present and is equal to the product of $B_z$ and $v_\phi$. Thus $B_z v_\phi$ must go to zero at the origin and since $v_\phi$ diverges due to conservation of angular momentum, $B_z$ must go to zero. In Appendix A we derived asymptotics of our solution close to the origin and found that stagnating solutions, shocked or not shocked, can only be obtained if $v_{\phi 0} < \chi^{1/2} v_{Az0}$.

\section{Empirical findings of stability}
Developing an analytical theory of stability of our self-similar Mag Noh problem requires significant
time and effort, so for the time being we employed Athena++ in cylindrical geometry to investigate stability in an empirical fashion. We modified boundary conditions in such a way as to perform a simulation with 5 different domains in azimuthal angle $\phi$, each domain covering $2\pi$ in angle.
In the first domain, we did not introduce perturbations, while four other domains were seeded with
a sinusoidal perturbation along $\phi$, corresponding to mode $m=\{1,2,4,8\}$ in one series of simulations and $m=\{3,5,6,10\}$ in the other. In the first series, the resolution was 64 grid points in each domain, and in the second -- 60 points. We chose resolution to be a common multiple of all $m$. Since we did not know the $m$ eigenmodes dependence in the radial direction we have chosen the perturbation to be confined within the shocked region of the solution, for this reason we introduced the perturbation, which was $10^{-3}$ in amplitude and present only in $B_\phi$ and $B_r$ (the latter to preserve divergence-free condition) starting with some initial time $t_i$ after which the shocked region has been already well-established. Had we chose to perturb multiple variables, this would complicate the matter since we do not know the phase between different vectors in eigenmode space.

Our first numerical experiments have demonstrated some unstable solutions, many of them grew very rapidly at the maximum $m$ possible (i.e. grid scale in $\phi$), seeded by truncation error. To counteract this we used a module to project the solution to the $\exp(-i m \phi)$ at each time step. Projection was also needed due to cross-contamination of modes. Even if we set the initial perturbation in energy to $10^{-6}$, due to rapid growth of instability it may not stay small to the end of the simulation. In this case it will produce other modes due to nonlinearity, which will grow on their own. The projection made sure that we always study a pure $\exp(-i m \phi)$ mode. The loss of perturbation energy due to the projection was small if the amplitude was small. If not, a sizable fraction of energy can be transferred to other modes and removed by the projection, which had a stabilizing effect. We should assume, therefore that the growth to large amplitudes is not accurate, see also discussion below.  
\begin{figure}
\begin{center}
\includegraphics[width=0.35\columnwidth]{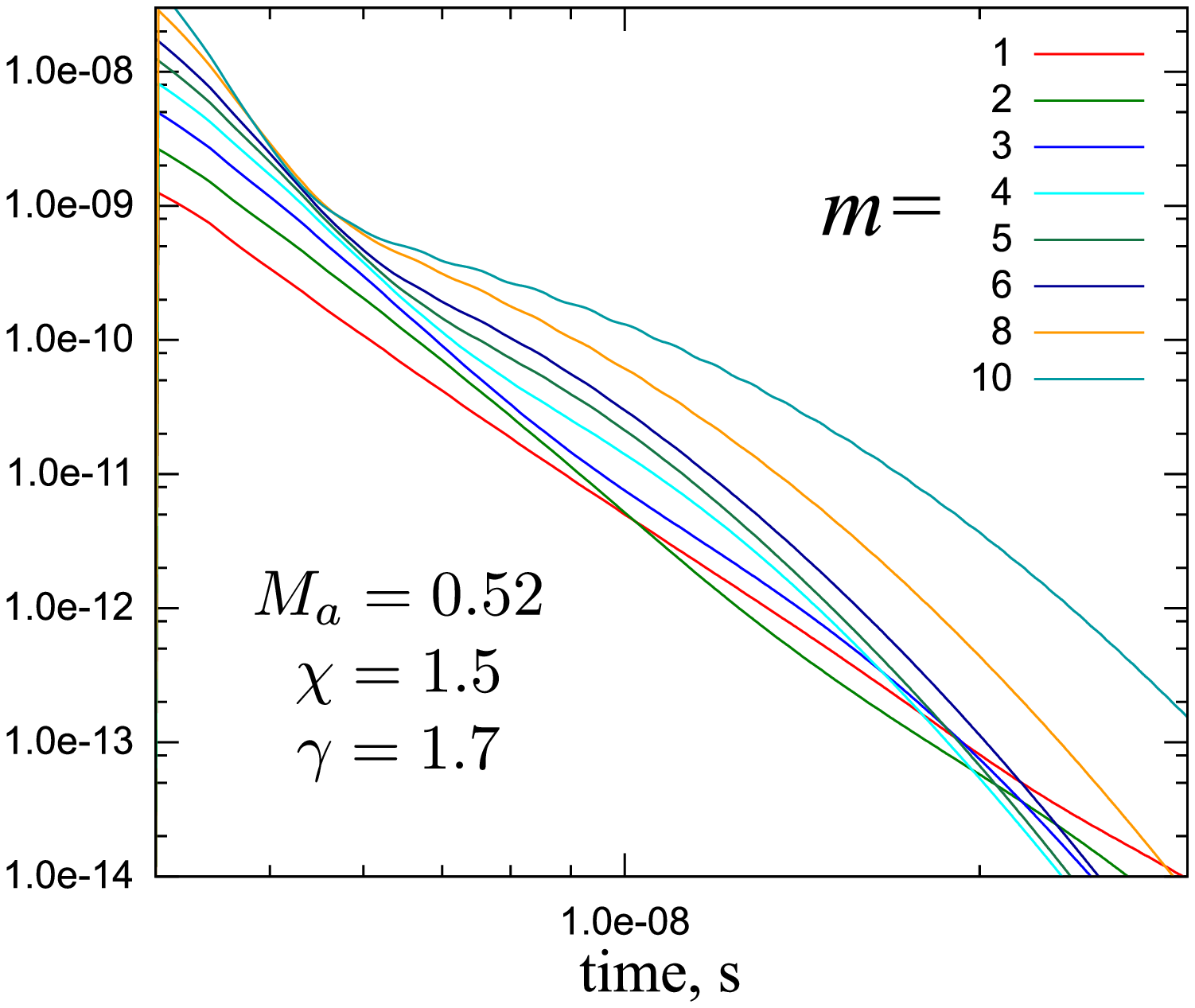}
\hfill
\includegraphics[width=0.35\columnwidth]{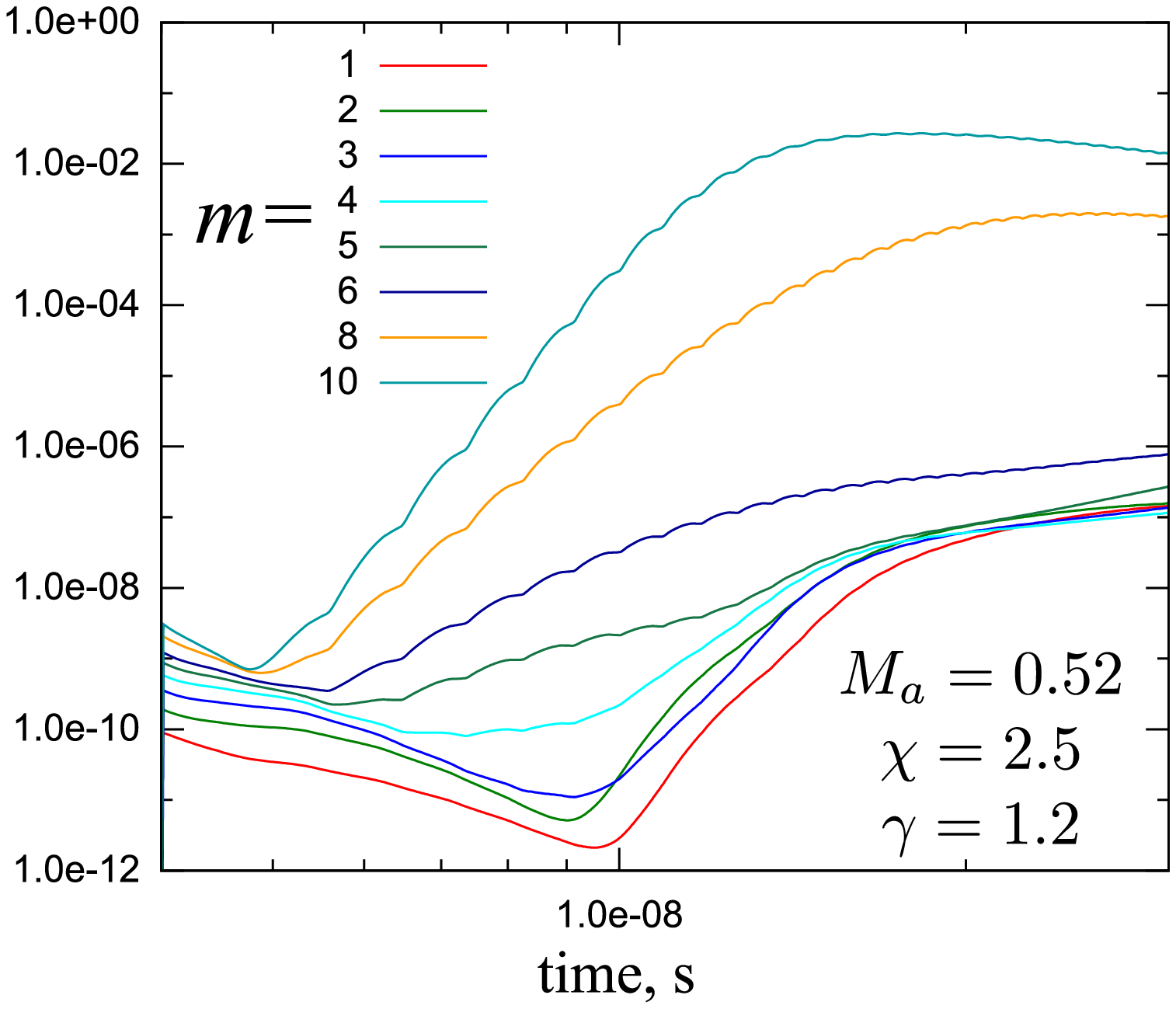}
\end{center}
\caption{Relative perturbation energies for different azimuthal mode numbers $m$ for one stable and one unstable solution of magnetized Noh.}
\label{st_and_unst}
\end{figure}

Figure~\ref{st_and_unst} demonstrates the ratio of the perturbation energy to the total energy of the shocked region as it evolves in time starting with $t_i=4\times 10^{-9}{\rm s}$ till the end of the simulation at $t_e=3\times 10^{-8}{\rm s}$. We had no way to guarantee that perturbation will stay small in all points, however, our projection module will eliminate any harmonics growing out of nonlinearity. Due to this the growth of the fastest-growing modes could have been somewhat suppressed. Despite their evolution was not accurate at later times we can still conclude that these quickly growing modes are unstable based on their evolution at earlier times. On Figure~\ref{st_and_unst} we have chosen two solutions from our parameter space, one of which we evaluate to be stable for all modes up to $m=10$ (except for 7 and 9, which we did not test), the other we judged to be unstable for all $m$ up to 10. To evaluate stability we compared maximum relative perturbation at earlier times, $0<t<5\times 10^{-9}{\rm s}$ and later times $6\times 10^{-9}{\rm s}<t<3\times 10^{-8}{\rm s}$. Needless to say, such evaluation should be viewed as preliminary, until the actual eigenmodes and eigenfrequencies are found.

\begin{figure}
\begin{center}
\includegraphics[width=0.55\columnwidth]{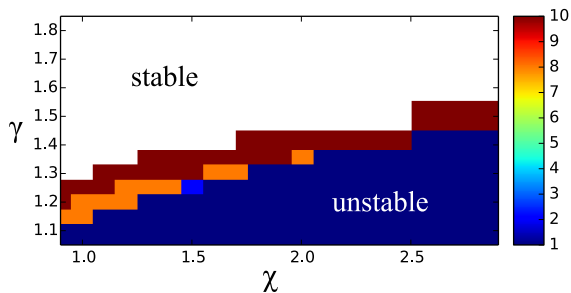}
\end{center}
\caption{Minimum unstable $m$ mode in the $\gamma,\chi$ place with $M_a=0.52$ and $B_z/B_\phi=1$.}
\label{m_min}
\end{figure}

We scanned our parameter space on the $\gamma,\chi$ plane, looking for stability. Figure~\ref{m_min} shows minimum unstable $m$, if exists. Primarily, self-similar solutions with low $\gamma$, high $\chi$, and high inflow velocity tend to be unstable. We also checked a few solutions with zero inflow velocity and found all of them stable.

We did not study stability analytically and did not find exact eigenmodes of azimuthal perturbation, however, the possibility that we are seeing numerical, rather than physical instability is rather remote. First of all, we are using Athena in cylindrical coordinates, so that the azimuthal perturbation is exactly along the coordinate axis. Numerical instability to azimuthal perturbations would require Athena to generate rather large fluxes of conservative quantities in the $\phi$ direction. Given that perturbations in $\phi$ direction are small and fluxes are generated based on the solution of the Riemann problem, this is unlikely. Secondly, conservative codes such as Athena are less likely to produce numerical instabilities. Thirdly, in many cases we see the instability to grow in a very well resolved $m=1$ mode, while numerical instabilities usually are generated on grid scale (large $m$). Fourthly, there is a clear dependence of the instability on physical parameters, such as $\gamma$, as shown on Fig.~\ref{m_min}. There is no reason to find such dependence in numerical codes, especially ones based on the solution of the Riemann problem.

\section{2D Cartesian simulations}
\begin{figure}
\begin{center}
\includegraphics[width=1.0\columnwidth]{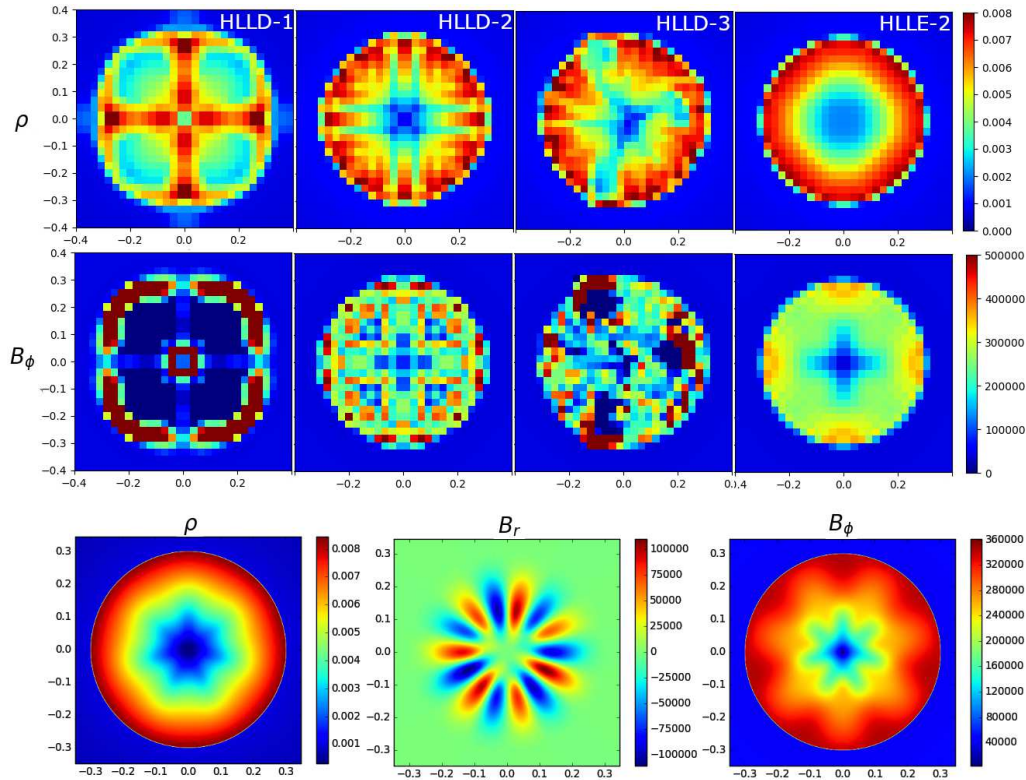}
\end{center}
\caption{2D Cartesian simulations of the same problem as presented on top of Fig.~\ref{two_solutions}. The top row shows density from four simulations using different numerical schemes: HLLD with 1st, 2nd and 3rd order spatial reconstruction and HLLE with 2nd order spatial reconstruction. The middle row shows $B_\phi$ for the same schemes as top row. The bottom row shows higher resolution HLLE 2nd order simulation, stabilized with grid-scale viscosity and resistivity. In addition, an m=7 perturbation with $10^{-2}$ amplitude was introduced in the $B_\phi$ and $B_r$ in the inflow.}
\label{2d_cart}
\end{figure}
In all the simulations above we used Athena in cylindrical geometry with first-order spatial reconstruction.
Sometimes the physics of the problem necessitates using Cartesian numerical simulations. For example, when an instability becomes nonlinear and transitions to turbulence, we prefer to study such turbulence on a uniform grid to avoid numerical effects associated with coordinate system's singular point. Despite extra errors, it is interesting to see how numerical codes in Cartesian case will deal with our self-similar solutions, including the unstable solutions. Cartesian coordinates will introduce discretization errors, C4-symmetric or periodic with $\pi/2$ in azimuthal angle, i.e. harmonics of $m=4$, in addition, there may be truncation errors, not necessarily symmetric. Figure~\ref{2d_cart} shows $\rho$ and $B_\phi$ on a 2D plain obtained in Athena++ simulation with different orders of spatial reconstruction. We initiated the problem with parameters $\chi=2.6, \gamma=1.1, v_A/V_s=0.84, v_0/V_s=20$, zero rotation. Earlier in this study we empirically found this solution unstable to all $m$ modes up to 10. We see from the figure that the instability seeded by C4-symmetric discretization error is dominant with first and second examples, but in the third, the instability was perhaps seeded by the truncation error as well. This is consistent with our cylindrical simulation where discretization error did not produce azimuthal perturbations, nevertheless, the instability grew from truncation error. In example four in Figure~\ref{2d_cart} we used a more diffusive solver, HLLE
and this solver produced broadly sensible, but inaccurate solution, especially for $B_\phi$. Finally, we found that stabilizing code with viscosity and magnetic diffusivity and going to higher resolution will provide accurate solutions. To test this we performed a convergence study with diffusivities, scaling as grid-scale, and observed convergence to a self-similar solution, albeit slow. On the bottom of Figure~\ref{2d_cart} we show similar simulation where we seeded initial $m=7$ perturbation in the inflow and observed that it is dominant almost everywhere, except near axis where the remaining $m=4$ perturbation
is still noticable. Our interest in simulating accurate unstable solutions is primarily to determine if one can obtain perturbed solution with the physically seeded perturbation. Collapsing Z-pinches are often unstable, however one often encounters numerical studies where the instability is seeded by numerical error. In this case the post-instability evolution and transition to turbulence is code-dependent and therefore meaningless. In order to properly study disruption due to instability one have to seed perturbations explicitly and provide an accurate solution.

\section{Discussion}
NRL Mag Noh is a family of self-similar solutions, meaning that the solution at any given time can be trivially rescaled to obtain a solution at some other time. This is especially convenient to estimate
the rate of convergence to an ideal solution in the case of fixed-grid MHD codes. Indeed, a solution at some time $t$ will have twice fewer grid points compared to the solution at $2t$, which makes it possible to compare accuracy between $t$ and $2t$, instead of running two different simulations, as it is traditionally done. 

This paper generalized the previously found family of self-similar implosions \cite{Velikovich2012} to the case with axial magnetic field and rotation. This makes our solution family five-parametric. Compared to this previous paper, which provided a reference solution in a tabulated form, we provide codes to generate solutions with arbitrary parameters, verify these solutions with an MHD code and test their stability. A physical initial value problem will require seven parameters: two components of velocity, two components of the magnetic field, density, self-similar index, and gas gamma. Two extra parameters correspond to the arbitrary choice of the shock velocity, as well as an arbitrary rescaling of magnetic field and density while keeping Alfv\'en speed constant. These two exact symmetries of ideal MHD are the only degeneracies of the problem and to the best of our knowledge, the variations of the five main parameters of our problem will always produce different solutions that can not be rescaled into each other. 

We expect the subset of the NRL Mag Noh problem with zero initial velocity to be especially useful to numerists due to the simplicity of its setup, for example, stationary setup is much easier in moving mesh codes and PIC codes. At the same time, zero initial velocity is grounded in the reality of magnetized implosions, where plasma is always initially at rest and is driven by magnetic tension. A subset of our solutions has a singular azimuthal current at the origin. Engineering such singularity in real implosions may help to create a hot spot in magnetically driven fusion experiments.   

In \cite{Velikovich2012} analytic expansion near origin was used to start integration. In this paper, we integrate from large to small radii and find a solution that stagnates at the origin by the bisection method. We found that understanding the appearance of the fast magnetosonic point is a critical element to achieve a regular solution. We use a sort of inverse logic to understand the property of magnetosonic point - if the regular solution exists, reducing inflow velocity will make it stagnate before origin, but will not produce a fast magnetosonic point. Therefore, if we encountered the fast magnetosonic point the inflow velocity is too high. Also, the inflow velocity is too high when the flow fails to stagnate at the origin. 
Bisecting between $|v_{r0}|$ too high and too low should always produce a regular stagnating solution. Interestingly, this argument does not prove that such a solution is unique and this interesting question will be considered elsewhere. An alternative method to produce a solution with a minimal $|v_{r0}|$, that we do not follow here, can be described as follows. In a range of $|v_{r0}|$ values there always is a self-similar solution with a shock, as long as we set a reflecting boundary at a certain radius $r_0$ (early stagnation). In order to obtain the solution that stagnates at origin, i.e., obtain $r_0=0$, we can slowly increase inflow velocity $|v_{r0}|$ so that $r_0$ goes to zero. In this case, by construction, we will obtain a unique solution with minimal $|v_{r0}|$.

We estimated the stability of our solutions by cylindrically symmetric simulations. It appears that there is a large parameter space where solutions are unstable to azimuthal perturbations. This can serve as a groundwork for future theoretical work to uncover the nature of this instability. Recent theoretical studies
of stability of the classic Noh problem, $B_z=B_\phi=0,\ \chi=0$ with non-ideal EoS \citep{Velikovich2016,Huete2021} developed criteria for D'yakov-Kontorovich instability. These can be applied
for the $B_z\neq 0,\ B_\phi=0$ case as well by treating the $B_z$ field as an additional fluid with $\gamma=2$. In this case the shock should be stable. Empirical study of a solution with $B_z=0,\ B_\phi\neq 0$
in \citet{Velikovich2012} did not find any instability, but we consider it feasible that some of these may be unstable. Our current empirical study of is in no way exhaustive and covers only a fraction of the parameter space.

Despite we do not fully understand the nature of the instability, it can be useful for theory and numerical research. While stable solutions traditionally played a role in estimating the accuracy of numerical codes, the unstable solutions are important to test the robustness of the codes and their ability to properly evaluate the development of the instability as well as its nonlinear evolution. It should be kept in mind that our stability setup keeps origin at rest, which may prevent some instabilities to be manifested. In addition, possible divergence of perturbations near origin may lead to a new continuum spectrum as in \cite{Sanz2016}.

Although we also produced solutions with finite initial pressure, we omit those, because they are not qualitatively different from solutions with zero initial pressure. In fact, to significantly change the shape of the flow, one needs to start with relatively high plasma beta, which is not a typical scenario in magnetically-driven HEDP plasmas. For completeness, our publicly released code can also produce solutions with finite initial pressure, making our solutions six-parametric.

In this paper we limited ourselves to self-similar solutions of ideal MHD equations that have a constant shock speed.
Our work, however, is not a formal mathematical exercise, it shows a complex nature of MHD stagnating solutions, rich in interesting physics. We hope that our family of MHD solutions will be used as a starting point in investigations of stagnation physics, shock physics, and stability. Comparing our ideal MHD solutions
with resistive-viscous cases, extended MHD, as well as PIC plasma may uncover more interesting physical effects, including induced rotation, instabilities, turbulence, particle acceleration, etc.  

\section{Data release}
\label{release}
The code needed to reproduce figures in this paper is available at 

\url{https://github.com/beresnyak/magnoh}

This release includes code to produce self-similar solutions, to estimate accuracy, to verify solutions with
Athena++, also it includes code for Athena++ to study stability to azimuthal perturbations. 

\section{Acknowledgement}
We would like to thank Dr. S. T. Zalesak, who provided us with his preliminary results on Mag Noh stability. We are grateful to Dr. C.~L.~Rousculp and Prof. C. E. Seyler for their insightful comments.
A. Beresnyak, A. L. Velikovich, J. L. Giuliani and A. Dasgupta were supported by the Department of Energy/National Nuclear Security Administration under the Interagency Agreement DE-NA0003278. A. Beresnyak was partially supported by the NASA Astrophysics Theory grant 80HQTR18T0065.

\section{Declaration of Interests}
The authors report no conflict of interest.

\appendix
\section{Asymptotic Expansions at small $\eta$}
Our equations (\ref{U_main})-(\ref{Az_main}) are written in terms of $\ln \eta$, so both infinity and zero are singular points.
We need to set the boundary condition for $U$ at $\eta=0$, however, and this boundary condition must be determined analytically.
We have stagnating flow at the origin, i.e., linear velocity gradient in $v_r$, so we can safely assume that $U$ is going to a negative constant as $\eta \to 0$ that we designate as $U_0$. Let us first determine dominant terms near zero in a simple qualitative fashion, expanding into more detailed calculations below.

First, from (\ref{Aphi_int}), $A_\phi \sim \eta^{-2}$, and subtracting (\ref{Aphi_main}) from (\ref{Az_main}),
near origin, we introduce positive $\alpha$ as:

\begin{equation}
\frac{d \ln A_z/A_\phi}{d \ln \eta} = \frac {2U_0}{1-U_0} = -\alpha. \label{AzAphi}
\end{equation}

From equations (\ref{S_int}-\ref{W_int}), (\ref{AzAphi}) we get

\begin{equation}
A_\phi \sim \eta^{-2}, \ \ A_z \sim W^2 \sim \eta^{-2-\alpha}, \ \ N \sim \eta^{2\chi-(\chi+1)\alpha}, \ \ S\sim \eta^{-2-(\gamma-1)\alpha}. \label{asymp_main}
\end{equation}

As we see, since $\alpha$ is positive, $v_\phi=\eta W \sim \eta^{-\alpha}$ is always singular on axis.
Also, sufficiently close to $\eta=0$, $A_\phi<A_z$, while $A_z$ and $W^2$ are of the same order. 
With $W^2/S \sim A_z/S \sim \eta^{\alpha(\gamma-2)}$ there are two cases in which either $S$ dominate,
or $A_z$ and $W^2$ dominate.

For $\gamma>2$, the dominant contribution to the main equation for $U$, (\ref{U_main}) will come from
terms with $S$. Neglecting non-essential terms in (\ref{U_main}) close to the origin, we can write $dU/d\ln \eta + 2\chi/\gamma+2U=0$, which only produces regular solution for $U$ if $U_0=-\chi/\gamma$.
In this case, $\alpha=2\chi/(\chi+\gamma)$ and the asymptotics of the physical quantities can be readily obtained as

\begin{equation}
N \sim \eta^{2\chi(\gamma-1)/(\chi+\gamma)}, \ \ H_\phi \sim \eta^{\chi(\gamma-1)/(\chi+\gamma)}, \ \ \ H_z \sim  \eta^{\chi(\gamma-2)/(\chi+\gamma)}, \ \ \ P \sim {\rm const}  \ \ ({\rm for} \ \gamma>2). \label{asympgg2}
\end{equation}

A more interesting and more physically relevant case $\gamma<2$ will see dominant contributions from $A_z$ and $W^2$ (note, that a special case with $A_z=W=0$ had been considered in \citet{Velikovich2012}). For
future reference we rewrite equation (\ref{U_main}) using integral (\ref{W_int}), designating the value of this integral
$\lambda=W^2 A_z^{-1}(1-U)^{-2}=(v_{\phi 0}/v_{Az0})^2$:

\begin{equation}
(U+S+A_\phi+A_z-1) \frac{d U}{d \ln \eta}+U(U-1)+2 S\left(U+\frac{\chi}{\gamma}\right)+(\chi+1)A_\phi
+(\chi+2U-\lambda(1-U)^2)A_z=0 \label{U_main_int}\\	
\end{equation}

In the absence of rotation, for $\lambda=0$ we get a regular solution for $U$ if and only if $U_0=-\chi/2$. From this $\alpha=2\chi/(\chi+2)$ and

\begin{equation}
N \sim \eta^{2\chi/(\chi+2)}, \ \ H_\phi \sim \eta^{\chi/(\chi+2)}, \ \ \ H_z \sim  {\rm const}, \ \ \ P \sim \eta^{(2-\gamma)2\chi/(\chi+2)}  \ \ ({\rm for} \ \gamma\leq 2) \label{asympgl2} .
\end{equation}

In the presence of rotation $U_0$ will be determined from the requirement that the coefficient for $A_z$ in (\ref{U_main_int}) vanishes, which is a quadratic equation $\chi+2U_0=\lambda(1-U_0)^2$. One solution is always
positive, which we can not use, the other is: 
\begin{equation}
U_0=\frac{\lambda+1-\sqrt{1+\lambda(\chi+2)}}{\lambda}
\end{equation}
In the limit of $\lambda \to 0$ we recover the above result $U_0=-\chi/2$. From the requirement that $U_0<0$
we obtain $\lambda<\chi$. This limits initial rotation which may still produce self-similar solution to
\begin{equation}
v_{\phi 0} < \chi^{1/2} v_{Az0}.
\end{equation}
This means that in the absence of $B_z$ we can only have rotating solutions for $\gamma>2$.

For the purpose of obtaining a self-similar solution with our method,
it is sufficient to have zeroth-order asymptotics for $U$. Let us, however, calculate a next correction to our asymptotics
above. To make the calculation compact, let us assume no rotation and $1<\gamma<2$. The calculation below can be repeated with rotation term, by adopting a different $U_0$ or with a different $\gamma$, which will change the ordering of terms. For our choice of $\gamma$ the dominant diverging term in (\ref{U_main_int}) will be $A_z$. Let us rewrite this equation rearranging terms of zeroth order and of lower order, assuming $\lambda=0$:
\begin{equation}
\frac{dU}{d\ln\eta}+2U+\chi+
\frac{S}{A_z}\left(\frac{dU}{d\ln\eta}+2U+\frac{2\chi}{\gamma}\right)+
\frac{A_\phi}{A_z}\left(\frac{dU}{d\ln\eta}+\chi+1\right)+
\frac{U-1}{A_z}\left(\frac{dU}{d\ln\eta}+U\right)
=0 \label{U_main3}
\end{equation}
We will ignore the last term and notice from (\ref{asymp_main}) that $S/A_z \sim \eta^\sigma$ and $A_\phi/A_z \sim \eta^\omega$ where
\begin{equation}
\sigma=\frac{2\chi(2-\gamma)}{\chi+2}, \ \ \ \omega=\frac{2\chi}{\chi+2}. \label{sigma}
\end{equation}
Consequently, the two lowest-order terms in our expansions will be either $\eta^\sigma$ and $\eta^{2\sigma}$  (when $3/2<\gamma<2$) or $\eta^\sigma$ and $\eta^\omega$ (when $1<\gamma<3/2$). Using equations (\ref{S_main}-\ref{Az_main}) we can write

\begin{equation}
\frac{d\ln(S/A_z)}{d\ln\eta}=-\frac{2-\gamma}{1-U}\left(\frac{dU}{d\ln\eta}+2U\right), \label{SAz}
\end{equation}

\begin{equation}
\frac{d\ln(A_\phi/A_z)}{d\ln\eta}=-\frac{2U}{1-U}. \label{SAphi}
\end{equation}

We define $\displaystyle \epsilon=\lim_{\eta\to 0} S/\eta^\sigma A_z$ and 
seek $S/A_z$ in the form of power series
\begin{equation}
\frac{S}{A_z}=\epsilon \eta^\sigma(1+\epsilon S_1\eta^\sigma+\epsilon^2 S_2\eta^{2\sigma}+\dots),
\end{equation}
And $U$ in the form of the power series
\begin{equation}
U=U_0+\epsilon U_1 \eta^\sigma + \epsilon^2 U_2 \eta^{2\sigma}+\dots.
\end{equation}
Neglecting terms with $A_\phi/A_z$ and $(U-1)/A_z$, equation (\ref{U_main3}) can be used to obtain $U_1$
and equation (\ref{SAz}) to obtain $S_1$:
\begin{equation}
U_1=-\frac{\chi(\chi+2)(2-\gamma)}{2\gamma(3\chi-\gamma\chi+2)},
\end{equation}
\begin{equation}
S_1=\frac{(2-\gamma)(2\chi-\gamma\chi+2)}{\gamma(3\chi-\gamma\chi+2)}.
\end{equation}
The process can be repeated to calculate next-order terms. In a similar manner, the terms with $\eta^\omega$ are calculated from (\ref{U_main3}) and (\ref{SAphi}) using $\displaystyle \delta=\lim_{\eta\to 0} A_\phi/\eta^\omega A_z$ and $U=U_0+\delta \bar U_1 \eta^\omega$ to obtain
\begin{equation}
\bar U_1=-\frac{(\chi+2)}{4}.
\end{equation}
Expansion of other quantities can be obtained from the expansion of $U$ and equations (\ref{ssim_rho}), (\ref{ssim_p}-\ref{ssim_hz}). For example, to the first order in $\epsilon$:
\begin{equation}
N \sim \eta^{\frac{2\chi}{\chi+2}}\left\{1- \frac{(4-\gamma)\chi+2}{\gamma[(3-\gamma)\chi+2]}\epsilon \eta^{\sigma}\right\},
\end{equation}
\begin{equation}
P \sim \eta^{\sigma}\left\{1- \frac{(3-\gamma)\chi+1}{\gamma[(3-\gamma)\chi+2]}\epsilon \eta^{\sigma}\right\},
\end{equation}
\begin{equation}
H_\phi \sim \eta^{\frac{\chi}{\chi+2}}\left\{1- \frac{(3-\gamma)\chi+1}{\gamma[(3-\gamma)\chi+2]}\epsilon \eta^{\sigma}\right\},
\end{equation}
\begin{equation}
H_z \sim 1-\frac{1}{\gamma} \epsilon \eta^{\sigma}.	
\end{equation}
We also see that $j_\phi \sim dH_z/d\eta \to 0$ at $\eta \to 0$ if $\sigma >1$. The same condition ensures that   $\nabla P$ vanishes on axis, so the pressure has a smooth maximum, not a sharp spike. From (\ref{sigma}), we find that this condition can be satisfied if $\gamma<3/2$, and $\chi$ is sufficiently large, $\chi>2/(3-2\gamma)$.
In the general case with rotation we have $\sigma = 2|U_0|(2-\gamma)/(|U_0|+1)$ and the condition $\sigma>1$
translates into $1/(3-2\gamma)<|U_0|<\chi/2$. We see that, when rotation is present, the regularity condition for  $j_\phi$ is not easier to meet because the parameters $\gamma<3/2$ and $\chi$ still need to satisfy $\chi>2/(3-2\gamma)$.

\section{Asymptotic Expansions at large $\eta$}
Using change of variables $\xi=1/\eta$ and introducing $\Phi$:
\def\Azinf{{v_{Az0}^2}}
\def\Apinf{{v_{A\phi 0}^2}}
\begin{equation}
A_z=\frac{\Azinf\Phi}{(1-U)^2}\xi^2, \label{Az_redef}
\end{equation}
such as $\Phi \to 1$ as $\xi \to 0$, we transform (\ref{Az_main}) to 
\begin{equation}
(U-1)\frac{d \ln \Phi}{d \ln \xi} -2U =0. \label{Phi_eq}
\end{equation}
Likewise the integral (\ref{W_int}) will be transformed into
\def\Winf{{v_{\phi 0}^2}} 
\begin{equation}
W^2=\Winf\Phi\xi^2. \label{Wint_redef}
\end{equation}
Substituting $S=0$, equations (\ref{Az_redef}), (\ref{Wint_redef} and (\ref{Aphi_int}) into 
the equation of motion (\ref{U_main}) we get
\begin{eqnarray}
&-&[(U-1)^3+(\Azinf\Phi+\Apinf)\xi^2]\frac{dU}{d \ln \xi} \nonumber \\
&+&\left\{[(2U+\chi)\Azinf-(1-U)^2\Winf]\Phi+(\chi+1)\Apinf\right\}\xi^2
+U(U-1)^3=0. \label{U_eq_redef}
\end{eqnarray}
Let us introduce asymptotic expansion at $\xi\to 0$:
\begin{equation}
	U=u_1\xi+u_2\xi^2+u_3\xi^3+\dots;
\end{equation}
\begin{equation}
	\Phi=1+\phi_1\xi+\phi_2\xi^2+\dots
\end{equation}
Here $u_1$ is our familiar $v_{r0}$. Substituting the expansion into the equation for $\Phi$, (\ref{Phi_eq}) to the second order in $\xi$ we obtain
\begin{equation}
-(\phi_1+2u_1)\xi+(-2\phi_2-2u_2+\phi_1^2+u_1\phi_1)\xi^2 =0.
\end{equation}
Equating both coefficients to zero we get
\begin{eqnarray}
	\phi_1&=&-2u_1;\\
	\phi_2&=&-u_2+u_1^2.
\end{eqnarray}
Substituting the expansion into the equation for $U$, (\ref{U_eq_redef}), we find first order term
to cancel out, leaving no constraint for $u_1$, which is our initial condition $v_{r0}$, the remainder being
\begin{eqnarray}
\left[(\chi+1)\Apinf+\chi\Azinf -\Winf+u_2\right]&\xi^2& \nonumber \\
+\left[u_1(\Azinf-\Apinf+2\Winf-3u_2)+(\chi\Azinf-\Winf)\phi_1+2u_3\right]&\xi^3&=0.	
\end{eqnarray}
From this we express all the coefficients via $v_{r0},v_{A\phi 0},v_{Az0}$ and $v_{\phi 0}$:
\begin{eqnarray}
u_2&=&-(\chi+1)\Apinf-\chi\Azinf +\Winf,\\
u_3&=&-\frac12 v_{r0} \left[ (3\chi+2)\Apinf+(\chi+1)\Azinf +\Winf\right],\\
\phi_1&=& -2 v_{r0},\label{phi_one}\\
\phi_2&=& (\chi+1)\Apinf+\chi\Azinf +v_{r0}^2-\Winf.\label{phi_two}
\end{eqnarray}
The asymptotic formula for self-similar velocity at large distance thereby is
\begin{equation}
\eta U=v_{r0}-\frac{(\chi+1)\Apinf+\chi\Azinf -\Winf}{\eta}
-\frac{\left[ (3\chi+2)\Apinf+(\chi+1)\Azinf +\Winf\right]v_{r0}}{2\eta^2}+\dots
\end{equation}
Without the axial magnetic field and rotation this reproduces the first three terms
of the asymptotic formula in \citet{Velikovich2012}. In case of zero initial velocity, 
$v_r \sim 1/\eta+{\rm const}/\eta^3+\dots$ at large distances. The asymptotic formula for the azimuthal velocity stems from substituting (\ref{phi_one}-\ref{phi_two}) into Eq. (\ref{Wint_redef}):
\begin{equation}
\eta W=v_{\phi 0} \left\{ 1-\frac1\eta v_{r0} 
+\frac1{2\eta^2}\left[(\chi+1)\Apinf+\chi\Azinf -\Winf\right]+\dots\right\}.
\end{equation}
The asymptotic expressions for other physical variables are derived by substituting the asymptotic expansion for $U$ into (\ref{ssim_rho}), (\ref{ssim_p}-\ref{ssim_hz}). These are:

\begin{eqnarray}
N&\sim&\eta^{2\chi}\left\{1-\frac{2\chi+1}\eta v_{r0} 
+\frac{\chi}{\eta^2}\left[(\chi+1)\Apinf+\chi\Azinf+(2\chi+1)v_{r0}^2 -\Winf\right]\right\},\\
P&\sim&\eta^{2\chi} \left\{1-\frac{2\chi+\gamma}\eta v_{r0}  
+\frac{\chi}{\eta^2}\left[(\chi+1)\Apinf+\chi\Azinf-\Winf\right] \right. \nonumber\\
&+&\left. \frac{(2\chi+\gamma)(2\chi+\gamma-1)}{2\eta^2}v_{r0}^2\right\}\\
H_\phi&\sim&\eta^{\chi}\left\{1-\frac{\chi}\eta v_{r0} 
+\frac{\chi-1}{2\eta^2}\left[(\chi+1)\Apinf+\chi\Azinf+\chi v_{r0}^2 -\Winf\right]\right\},\\
H_z&\sim&\eta^{\chi}\left\{1-\frac{\chi+1}\eta v_{r0} 
  +\frac{\chi}{2\eta^2}\left[(\chi+1)\Apinf+\chi\Azinf+(\chi+1)v_{r0}^2 -\Winf\right]\right\}.
\end{eqnarray}

In the case of the plasma initially at rest, the asymptotic terms with $1/\eta$ and $1/\eta^3$ are absent,
leaving only the terms with $1/\eta^2$, $1/\eta^4$, etc.

\begin{figure}
\begin{center}
\includegraphics[width=0.8\columnwidth]{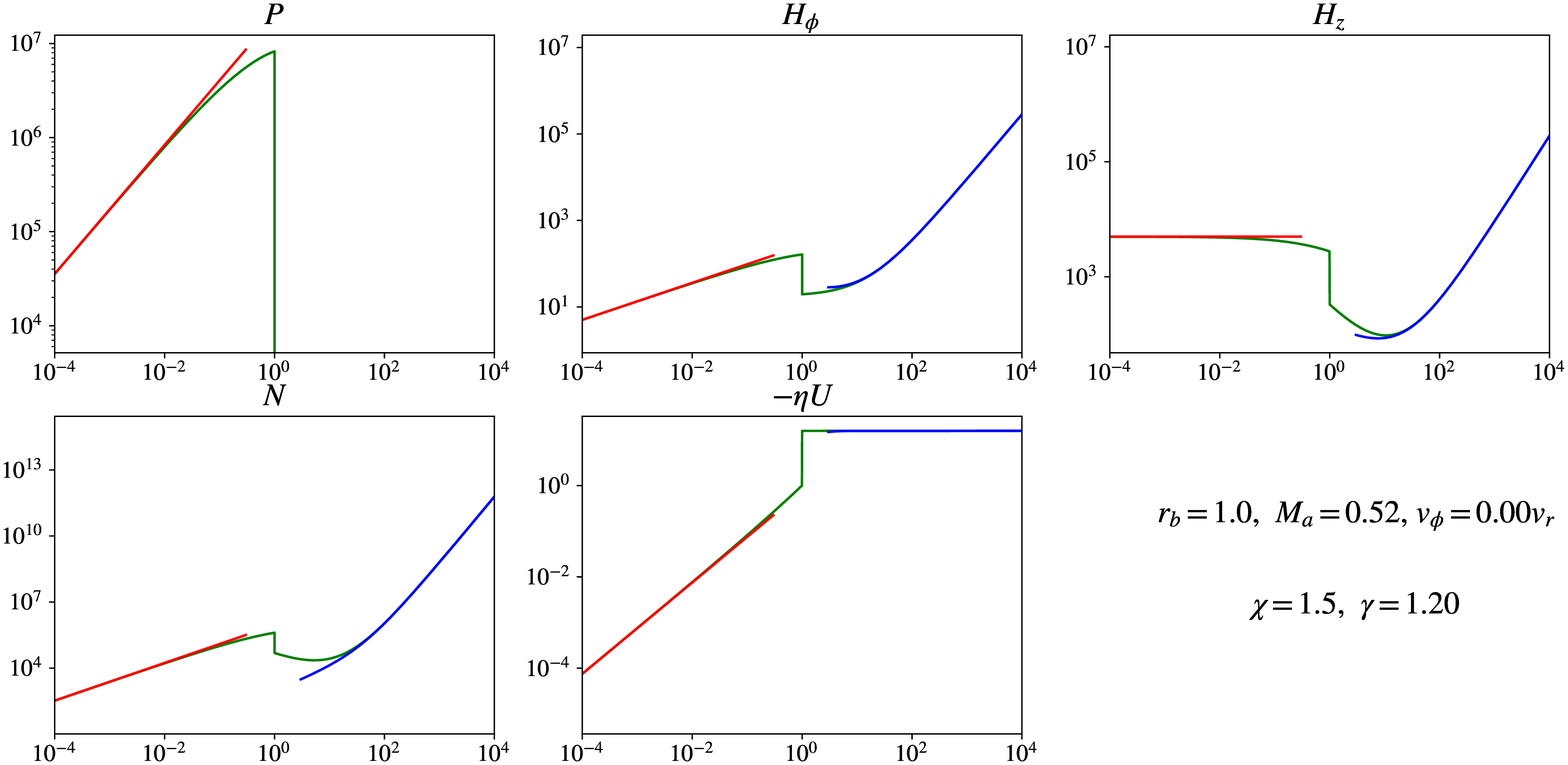}
\end{center}
\caption{Self-similar solution (green) vs asymptotic expression for small (red) and large (blue) $\eta$. }
\label{asympt}
\end{figure}

In Fig.~\ref{asympt} we showed the asymptotic limits plotted over our self-similar solution.

\par
\ \ 
\par

\def\apj{{\rm Astrophys. J.}}           
\def\apjl{{\rm ApJ }}          
\def\apjs{{\rm ApJ }}          
\def\grl{{\rm GRL }}
\def\aap{{\rm A\&A } }
\def\jgr{{\rm JGR}}
\def\mnras{{\rm MNRAS } }
\def\physrep{{\rm Phys. Rep. } }               
\def\prl{{\rm Phys. Rev. Lett.}} 
\def\pre{{\rm Phys. Rev. E}} 
\def\araa{{\rm Ann. Rep. A\&A } }
\def\prd{{\rm Phys. Rev. D}} 
\def\pra{{\rm Phys. Rev. A}} 
\def\ssr{{\rm SSR}}
\def\planss{{\rm Plan. Space Sciences}}
\def\apss{{\rm Astrophysics and Space Science}}
\def\nat{{\rm Nature}}
\def\jcap{{\rm JCAP}}
\def\memsai{{\rm MEMSAI}}
\def\lt{{<}}
\def\aapr{{\rm AAPR}}
\def\solphys{{\rm SolPhys}}

\bibliographystyle{jfm}
\bibliography{dpf}

\end{document}